# The Nonlinear Quantum Gauge Theory–Superrelativity

P. Leifer

Mortimer and Raymond Sackler Institute of Advanced Studies
Tel-Aviv University, Tel-Aviv 69978, Israel

**Abstract**

A new type of nonlinear gauge quantum theory (superrelativity) has been proposed [1, 2, 3, 4, 5, 6]. Such theory demands a radical reconstruction of the quantum field conception and spacetime structure, and this paves presumably way to the comprehension of the quantum nature of inertia.

## 1 Introduction

The superrelativity differs from ordinary gauge theories in the sense that the affine connection

$$\Gamma^i_{kl} = -2 \frac{\delta^i_k \Pi^{l*} + \delta^i_l \Pi^{k*}}{R^2 + \sum_s^N |\Pi^s|^2}, \tag{1.1}$$

as well as in general relativity, is constructed from first derivatives of the Fubini-Study metric tensor

$$G_{ik*} = R^2 \frac{(\sum_{s=1}^N |\Pi^s|^2 + R^2)\delta_{ik} - \Pi^{*i}\Pi^k}{(\sum_{s=1}^N |\Pi^s|^2 + R^2)^2} \tag{1.2}$$

of the projective Hilbert space of the pure quantum states CP(N). That is we have not merely analogy with general relativity (GR) but presumably this construction paves the way to the unification of GR and quantum theory. The theory demands a radical reconstruction of the quantum field conception and spacetime structure which should be done in order to overcome the stumbling block in foundations of quantum theory.

The principle of superrelativity [1, 2, 3] states as follows:

**The general unitary motion of the pure quantum states may be locally reduced to the geodesic motion in the projective Hilbert space by the introduction of a gauge (compensation) field which arises from the local**

**change of a functional frame in the original Hilbert state space**. Since the geodesic flow induced by transformations from the coset $SU(N+1)/S[U(1) \times U(N)]$, in distinction from the norm-preserving gauge transformations of Doebner-Goldin [7] **one has an observable deformation of the quantum state**. The gauge field is a physical evidence of this deformation. That is one has a "superequivalence" of the compensation field and some class of unitary motions of functional basis in Hilbert space as well as in GR one has the local equivalence of gravity and an accelerated motion of a frame in spacetime. **This gauge field is a function of the sectional curvature $\kappa = R^{-2}$ of the projective Hilbert space CP(N) which plays the role of the universal single coupling constant** [1, 2, 3].

## 2 Coset as Universal Quantum Interactions

We have put at the basis of our theory the fact that in *all interactions of quantum ("elementary") particles there is a conservation law of the electric charge.* Then unitary group SU(N+1) takes the place of *dynamical group* because in the general case $g|\Psi> \neq |\Psi>$ and, of course, $g^{-1}Hg \neq H$. There is the group of isotropy $H = U(1)_{el} \times U(N)$ of a fixed pure quantum state $e^{i\gamma}|\Psi>$. That is transformations which effectively act on this ray lie in the coset $G/H = SU(N+1)/S[U(1) \times U(N)]$. It is clear that deformations of the pure quantum state are due to some physical interaction; the effect of the interaction has the geometric structure of a coset, i.e. the structure of the CP(N): $G/H = SU(N+1)/S[U(1)_{el} \times U(N)] = CP(N)$. This paves the way to the invariant study of the spontaneously broken unitary symmetry [6]. This statement has a general character and does not depend on particular properties of the pure quantum state. **The reason for the change of motion of material point is an existence of a force. The reason for the change of a pure quantum state is an interaction which may be modeled by unitary transformations from the coset G/H. The reaction of a material point is an acceleration. The reaction of a pure quantum state is its deformation whose geometry is in fact the geometry of the coset transformations $SU(N+1)/S[U(1) \times U(N)]$. That is a universal quantum interactions at the fundamental level "live" in Hilbert projective state space CP(N)**.

## 3 Is all matter "dissolved" in CP(N) geometry?

When we build our theories we usualy begin with some fields on spacetime manifold. Thus, we think that it is the most realistic approach to the achievement of an agreement between theory and experiment. But we should remember that the measurement of the spacetime position of the "event" is **only the registration of the fact of some quantum transition in a quantum state space** [1, 2, 3, 4].

Therefore we can assume that from the physical point of view just the dynamics of a quantum transition is interesting for us. Hence, it is more natural to deal with functions on quantum state space than with the spacetime functions (fields). That is, instead of a material point and functions of its coordinates in spacetime, one should use functions of the coordinates of a quantum state. Otherwise we obtain singular results because a pointwise (in spacetime) dynamical variables contain singular functions. In fact, the all known variants of the regularization procedure are the processes of a "delocalization" of the pointwise dynamical variables. **Thus, in our case we search functions of "quantum transition" in the space of the "internal" degrees of freedom not functions of coordinates of the "event" in spacetime**. Dynamical (habitual) spacetime arises only as a manifold of the "centrum of mass" of quantum "droplets"–the soliton-like solutions of some nonlinear wave equations. As a matter of fact they are "bundles of energy and momentum" [9].

In order to derive an equation which presumably may to have soliton-like solution we apply the **"descent-geodesic variation-lift-unitary rotation"** –ansatz to the special form of the ordinary Klein-Gordon equation [1, 2, 3, 4]. **It should play the role of a model of non-local quantum particles in the framework of the causal approach to quantum theory**. For the global "shaping" of field configuration, I will choose the Lagrangian which depends only on the "radial" variable $\rho$, i.e. $\Phi = \Phi(\rho)$, where $\rho^2 = x_\mu x^\mu$. Then the Lagrangian density may be written as

$$\mathcal{L} = \frac{1}{2}\Phi^*_\mu \Phi^\mu - U(\Phi(\rho)) = \frac{1}{2}\left|\frac{d\Phi}{d\rho}\right|^2 - U(\Phi(\rho)), \tag{3.1}$$

where we have assumed the general form for the effective self-interaction term $U(\Phi(\rho))$. The equation of motion of the scalar field acquires the form of the ordinary differential (nonlinear in general) equation

$$\frac{d^2\Phi^*}{d\rho^2} + (3/\rho)\frac{d\Phi^*}{d\rho} + 2\frac{\partial U(\Phi(\rho))}{\partial \Phi} = 0. \tag{3.2}$$

If the potential $U(\Phi(\rho))$ has the form $U(\Phi(\rho)) = \frac{1}{2}(mc/\hbar)^2 \Phi^*\Phi = \frac{1}{2}\alpha^2|\Phi|^2$, then (3.2) is the linear Lommel equation

$$\frac{d^2\Phi^*}{d\rho^2} + (3/\rho)\frac{d\Phi^*}{d\rho} + \alpha^2 \Phi^* = 0, \tag{3.3}$$

for which a solution is expressed in the Bessel function $\Phi = \rho^{-1} J_{-1}(\alpha\rho)$. **However a "deformation" of these solutions into solutions of some effectively non-linear Klein-Gordon equation by the geodesic flow is interesting for our purpose**. If we intend to accept the hypothesis of the unified field mass seriously, one ought to take into account a visible fact that the classical radius of the electron is closed to the classical radius of the proton *in the reference Minkowski spacetime.*

This leads toward the corresponding definition of the strong coupling constant. **But in the framework of a unified field theory should be exist a single universal coupling constant. We suspect that in nonlinear approach it should be closed to $\alpha$ [1, 2, 3]**. Then if we choose the the classical radius of the electron $r_0 = \frac{e^2}{mc^2}$ as the unit of the Minkowski spacetime distance scale $\rho = xr_0$, then $(mc/\hbar)^2$ in (3.3) becomes the fine structure constant $\alpha = \frac{e^2}{\hbar c}$. **Hereafter this constant plays the role of the some fundamental "vacuum field mass", but the real field mass and dynamical size of effective area action of the unified field arise in dynamical manner from the oscillations near the bottom of the "potential of the projective connection (1.1)" as we will see soon**. Let us suppose $y = (\rho/r_0)^2$. One can represent a solution of (3.3) in the $y$-variable as a Fourier series

$$|\Phi> = x^{-1} J_{-1}(\alpha x) = \sum_{k=0}^{\infty} \Phi^k |k, y>, \tag{3.4}$$

where

$$\Phi^m = \frac{-1}{2^m m! \sqrt{\pi}} \int_{-\infty}^{\infty} dy \exp(-\frac{y^2}{2}) H_m(y) y^{-1/2} J_1(\alpha y^{1/2}) \tag{3.5}$$

and $|n, y> = (2^n n! \sqrt{\pi})^{-1/2} \exp(-y^2/2) H_n(y)$ is complete set of orthogonal Hermitian functions on the interval $(-\infty, \infty)$. Our geodesic flow acts on these Fourier components. It is easily seen that for a vector of this form, geodesic flow is generated by a general linear combination of "creation-annihilation" operators

$$\hat{B} = \begin{pmatrix} 0 & f^{1*} & f^{2*} & . & . & . & f^{N*} \\ f^1 & 0 & 0 & . & . & . & 0 \\ f^2 & 0 & 0 & . & . & . & 0 \\ . & . & . & . & . & . & . \\ . & . & . & . & . & . & . \\ . & . & . & . & . & . & . \\ f^N & 0 & 0 & . & . & . & 0 \end{pmatrix}. \tag{3.6}$$

The flow is then given by the unitary matrix $\hat{T}(\tau, g) = \exp(i\tau \hat{B}) =$

$$\begin{pmatrix} \cos\Theta & \frac{-f^{1*}}{g}\sin\Theta & . & . & . & \frac{-f^{N*}}{g}\sin\Theta \\ \frac{f^1}{g}\sin\Theta & 1 + [\frac{|f^1|}{g}]^2(\cos\Theta - 1) & . & . & . & \frac{f^1 f^{N*}}{g^2}(\cos\Theta - 1) \\ . & . & . & . & . & . \\ . & . & . & . & . & . \\ . & . & . & . & . & . \\ \frac{f^N}{g}\sin\Theta & \frac{f^{1*} f^N}{g^2}(\cos\Theta - 1) & . & . & . & 1 + [\frac{|f^N|}{g}]^2(\cos\Theta - 1) \end{pmatrix}, \tag{3.7}$$

where $g = \sqrt{\sum_{k=1}^{N} |f^k|^2}$, $\Theta = g\tau$ [1, 2, 3]. The form of the periodic geodesic "deformation" of the initial solution of (3.3) is represented by the formula

$$|\Psi(\tau, g, y)> = \sum_{m,n=0}^{\infty} \Phi^m [\hat{T}^{-1}(\tau, g)]_m^n |n, y>. \tag{3.8}$$

*Note, this deformation does not lead to so-called zeroth modes.*

The state vector (3.4) inherits the geomeric structure of the $CP(\infty)$ due to the "descent-geodesic variation-lift-unitary rotation" –ansatz [1, 2, 3], and the perturbed Lagrangian is as follows:

$$\mathcal{L}' = \mathcal{L}(\Psi + \Delta\Psi, \Psi^* + \Delta\Psi^*)$$
$$= \frac{1}{2}\frac{d(\Psi + d\Delta\Psi)}{d\rho}\frac{d(\Psi + d\Delta\Psi)^*}{d\rho} - \frac{1}{2}\alpha^2(\Psi + \Delta\Psi)(\Psi + \Delta\Psi)^*, \quad (3.9)$$

where $\Delta\Psi^i = -\Psi^0\Gamma^i_{km}\xi^k d\Pi^m\tau$. Then one has a nonlinear Klein-Gordon equation in the Lommel form plus the nonlinear terms

$$\frac{d^2(\Psi + \Delta\Psi)^*}{d\rho^2} + (3/\rho)\frac{d(\Psi + \Delta\Psi)^*}{d\rho} + Y(\Delta\Psi, \rho)$$
$$+\alpha^2(\Psi^* + \Delta\Psi^* + \Psi^*\frac{\partial\Delta\Psi}{\partial\Psi}) = 0, \quad (3.10)$$

where $Y(\Delta\Psi, \rho)$ denotes terms which arise under the variation of derivations of $\Delta\Psi$, and where for enough small $\tau$

$$\Delta\Psi^i = -g\Psi^0\tau^2(1 + \frac{|\Psi^0|^2}{R^2})^{-1/2}\Gamma^i_{km}\xi^k\Psi^m. \quad (3.11)$$

If the radius $R$ of the sectional curvature $\kappa = 1/R^2$ of the projective Hilbert space goes to infinity, one obtains the ordinary Klein-Gordon equation.

At first sight our theory looks like scalar electrodynaics where no room for the spin degrees of freedom. But we should take into account the possibility of an identification of the "soliton" of the nonlinear Klein-Gordon equation (3.10) and the fermion of the Dirac theory. It connected with the transformation of the "isospin" of our scalar multiplet into the spin. Besides this one should remember that gauge group of our theory coinsides with the isotropy group $H = U(1) \times U(N)$ of the local ground state (vacuum) and CP(N) has nontrivial topological structure. (Let us recall that Coleman [10] had proved the equivalence of the quantum S-G soltion with the fermion of the Thirring model [11]).

# 4  A Geometrical "Second" Quantization

From the formal point of view the modern quantum field theory is based on operator-value functonals which defined on functions over spacetime manifold. Such theory cames from the second quantization scheme of Dirac [8]. In the framework of statistical properties it is very powerful method but on the fundamental level it has not a dynamical contents and, therefore, is limited. Besides that the second quantization method brings a mesh of pure mathematical problems such as multiplication of the singular functions. I think that these are merely artefacts as a consequence of our wrong point of view on two fundamental things: **spacetime manifold and quantum field amplitudes**. Here we make accent on the geometric aspect of the second quantization which leads toward the **natural one-to-one correspondence between Fourier amplitudes of the single scalar field and linear combination of the "creation-annihilation" operators of self-interaction "quants" between these amplitudes**.

Let us assume that we wish to understand how one can link in the functional space the vacuum state

$$|\Psi_0>= \begin{pmatrix} e^{i\omega(\Psi)}\sqrt{\sum_{a=0}^{N}|\Psi^a|^2} \\ 0 \\ \cdot \\ \cdot \\ \cdot \\ 0 \end{pmatrix} = Re^{i\omega(\Psi)}\begin{pmatrix} 1 \\ 0 \\ \cdot \\ \cdot \\ \cdot \\ 0 \end{pmatrix}, \qquad (4.1)$$

and, say, solution of the Lommel equation (3.3). Only one path from the indefinite number of pathes is interesting for us–just the geodesic between them in the Hilbert projective space CP(N). In order to find the elements $f^i$ of the generator of the geodesic flow for deformation of the vacuum state toward the solution (3.4) one should note that if the result of the periodic geodesic "deformation" of the initial solution of (3.8) is not so far from (4.1), then one can span them by an unique geodesic $(1,0,...,0,...)\hat{T}(\tau, g) = R^{-1}(\Phi^0, \Phi^1, ..., \Phi^N, ...)$. It may be shown that $cos\Theta = |\Phi^0|/R$, $|f^i| = g|\Phi^i|(R^2 - |\Phi^0|^2)^{-1/2}$ and $\arg f^i = \arg \Phi^i$ (up to the general phase). Thus the "rate" and direction of the transformation of the vacuum vector into the solution of the Lommel equation is determined by the matrix $\hat{P} = \hat{B}(\Phi)$ (3.6). But in order to avoid a worry about the structure of the operators of "creation-annihilation" for arbitrary initial state vector not just vacuum form (4.1), it is much more simpler to use local coordinates $\{\Pi\}$ in which for every N one can express and choose local generators with commutation relations from AlgSU(N+1) [1]. That is, we now refer to the term "local" as a fact of a dependence on the local coordinates in the $CP(N)$. We should find the relationship between the linear representations of $SU(N + 1)$ group by an "polarization operator" $\hat{P} \in AlgSU(N+1)$ which does not depend on the state of the quantum system and the nonlinear representation (realization) of the group symmetry in which the infinitesimal operator of the transformation depends on the state. In the linear representation of the action of $SU(N + 1)$ we have

$$|\Psi(\epsilon)>= \exp(-i\epsilon\hat{P})|\Psi>. \qquad (4.2)$$

For a full description of a group dynamics by pure quantum states, we shall use coherent states in $CP(N)$. Let us assume that $\hat{P}_\sigma$ is one of the $1 \leq \sigma \leq (N+1)^2 - 1$ directions in the group manifold. Then

$$D_\sigma(\hat{P}) = \Phi^i_\sigma(\pi, P)\frac{\delta}{\delta \pi^i} + \Phi^{i*}_\sigma(\pi, P)\frac{\delta}{\delta \pi^{i*}}, \qquad (4.3)$$

where

$$\Phi^i_\sigma(\pi; P_\sigma) = \lim_{\epsilon \to 0} \epsilon^{-1}\left\{\frac{[\exp(i\epsilon P_\sigma)]^i_m \Psi^m}{[\exp(i\epsilon P_\sigma)]^k_m \Psi^m} - \frac{\Psi^i}{\Psi^k}\right\} = \lim_{\epsilon \to 0} \epsilon^{-1}\{\pi^i(\epsilon P_\sigma) - \pi^i\} \qquad (4.4)$$

are the local (in CP(N)) state-dependent components of generators of the SU(N+1) group, which are studied in [1, 2, 3]. So in the general case for group transformations of more than one parameter we have a vector field for the group action on CP(N) by some set of dynamical variables $\hat{P}_1, ..., \hat{P}_\sigma, ..., \hat{P}_{(N+1)^2-1}$, as

$$V_P(\pi, \pi^*) = \sum_\sigma [\Phi^i_\sigma(\pi, P)\frac{\delta}{\delta \pi^i} + \Phi^{i*}_\sigma(\pi, P)\frac{\delta}{\delta \pi^{i*}}]\epsilon^\sigma. \qquad (4.5)$$

Then the differential of some differentiable function $F(\pi, \pi^*)$ is

$$\delta_P F(\pi, \pi^*) = D_\sigma(\hat{P})F(\pi, \pi^*)\epsilon^\sigma, \qquad (4.6)$$

and, in particular, we have

$$\delta_P \pi^i = \Phi^i_\sigma(\pi, P)\epsilon^\sigma, \delta_P \pi^{i*} = \Phi^{i*}_\sigma(\pi, P)\epsilon^\sigma. \qquad (4.7)$$

For example, realizing rotations $\hat{s}_x, \hat{s}_y, \hat{s}_z$ from $AlgSU(2)$, one has

$$D_x(s) = -\frac{\hbar}{2}[[1-\pi^2]\frac{\delta}{\delta \pi} - [1-\pi^{*2}]\frac{\delta}{\delta \pi^*}],$$
$$D_y(s) = \frac{\hbar}{2}[[1+\pi^2]\frac{\delta}{\delta \pi} + [1+\pi^{*2}]\frac{\delta}{\delta \pi^*}],$$
$$D_z(s) = \hbar[-\pi\frac{\delta}{\delta \pi} + \pi^*\frac{\delta}{\delta \pi^*}]. \qquad (4.8)$$

Then, we have well known commutation relations

$$[D_\mu(s), D_\nu(s)]_- = -i\hbar\epsilon_{\mu\nu\sigma}D_\sigma(s). \qquad (4.9)$$

For a three-level system, the realization of a dynamical $SU(3)$ group symmetry is provided by an 8-dimensional local vector field [1, 2, 3], where $\hat{\lambda}_1, ..., \hat{\lambda}_8$ are the Gell-Mann matrices, i.e.

$$D_1(\lambda) = i\frac{\hbar}{2}[[-1+(\pi^1)^2]\frac{\delta}{\delta \pi^1} + \pi^1\pi^2\frac{\delta}{\delta \pi^2} + [-1+(\pi^{1*})^2]\frac{\delta}{\delta \pi^{1*}} + \pi^{1*}\pi^{2*}\frac{\delta}{\delta \pi^{2*}}],$$
$$D_2(\lambda) = i\frac{\hbar}{2}[[1+(\pi^1)^2]\frac{\delta}{\delta \pi^1} + \pi^1\pi^2\frac{\delta}{\delta \pi^2} - [1+(\pi^{1*})^2]\frac{\delta}{\delta \pi^{1*}} - \pi^{1*}\pi^{2*}\frac{\delta}{\delta \pi^{2*}}],$$

$$D_3(\lambda) = -\frac{\hbar}{2}[\pi^2\frac{\delta}{\delta\pi^2} + \pi^{2*}\frac{\delta}{\delta\pi^{2*}}],$$
$$D_4(\lambda) = \frac{\hbar}{2}[[-1+(\pi^2)^2]\frac{\delta}{\delta\pi^2} + \pi^1\pi^2\frac{\delta}{\delta\pi^1} + [-1+(\pi^{2*})^2]\frac{\delta}{\delta\pi^{2*}} + \pi^{1*}\pi^{2*}\frac{\delta}{\delta\pi^{1*}}],$$
$$D_5(\lambda) = \frac{\hbar}{2}[[1+(\pi^2)^2]\frac{\delta}{\delta\pi^2} + \pi^1\pi^2\frac{\delta}{\delta\pi^1} - [1+(\pi^{2*})^2]\frac{\delta}{\delta\pi^{2*}} - \pi^{1*}\pi^{2*}\frac{\delta}{\delta\pi^{1*}}],$$
$$D_6(\lambda) = -\frac{\hbar}{2}[\pi^2\frac{\delta}{\delta\pi^1} + \pi^1\frac{\delta}{\delta\pi^2} - \pi^{2*}\frac{\delta}{\delta\pi^{1*}} - \pi^{1*}\frac{\delta}{\delta\pi^{2*}}],$$
$$D_7(\lambda) = -\frac{\hbar}{2}[\pi^2\frac{\delta}{\delta\pi^1} - \pi^1\frac{\delta}{\delta\pi^2} - \pi^{2*}\frac{\delta}{\delta\pi^{1*}} + \pi^{1*}\frac{\delta}{\delta\pi^{2*}}],$$
$$D_8(\lambda) = 3\hbar[\pi^2\frac{\delta}{\delta\pi^2} - \pi^{2*}\frac{\delta}{\delta\pi^{2*}}]. \quad (4.10)$$

In each of $N+1$ charts of the local coordinates these vector fields might be distinguished to two parts : Goldstone subspace $B$ and Higgs subspace $H$ with commutation relations of $Z_2$ -graded algebra $AlgSU(N)$: $[\hat{H},\hat{H}]_- \subset H, [\hat{H},\hat{B}]_- \subset B, [\hat{B},\hat{B}]_- \subset H$ like ordinary (state-independent) elements of $AlgSU(N+1)$.

**One should not confuse these "creation-annihilation" local operators with creation-annihilation operators of real quants which move in spacetime!** They are rather creation-annihilation operators of "quants of polarization" which shape the "droplet" as a carrier of these polarizations. But they pave the way to the introduction of a dynamical picture of the quantum behavior.

## 5 Goldstone and Higgs Geometric Mechanism

It has been shown [1] that in our original Hilbert space $\mathcal{H}$ the term $|dv(y)>$ arises as an additional rate of a change of the state vector $|\Psi>$

$$|dv(y)> = -\frac{i}{\hbar}d\hat{P}|\Psi(y)> = \sum_{s=1}^{\infty}\frac{d}{d\tau}(\Psi^0\Delta\Pi^s)\Big|_0 |s,y> \quad (5.1)$$

where $\Delta\Pi^i = -\Gamma^i_{km}\xi^k d\Pi^m \tau$. Then $\Delta U = \frac{\delta U}{\delta\Pi^i}d\Pi^i + \frac{\delta U}{\delta\Pi^{*i}}d\Pi^{*i}$ where $\frac{\delta U}{\delta\Pi^i} = -\hbar\Gamma^s_{ik}\xi^k|s,y>$ may be treated as an "instantaneous" (in the sense of the local in $CP(\infty)$ unitary rotation of the functional basis (frame)) self-interacting potential of the scalar configuration associated with the infinitesimal gauge transformation of the local frame with coefficients (1.1). Then dynamical spacetime energy distribution [4] of the oscillation modes is described by the formula

$$\frac{\delta^2 U}{\delta\Pi^i \delta\Pi^{*j}} = -\hbar\{\Gamma^s_{ik}\frac{\delta\xi^k}{\delta\Pi^{*j}} + \frac{\delta\Gamma^s_{ik}}{\delta\Pi^{*j}}\xi^k\}|s,y> = -\hbar\{\Gamma^s_{ik}\frac{\delta\xi^k}{\delta\Pi^{*j}} + R^s_{kj*i}\xi^k\}|s,y>, \quad (5.2)$$

where $R^s_{kj*i}$ is Riemannian tensor in $CP(\infty)$. That is the condition of the self-preservation of the droplet in visinity of solution of the ordinary Klein-Gordon equation presumably may be connected with the stability of geodesics (the sectional curavature $\kappa = R^{-2}$ of the projective Hilbert space is ever positive). Then the radiation of the gauge (compensation) field arises due to the renormalization of dynamical variables and "rotation" of the ellipsoid of polarization.

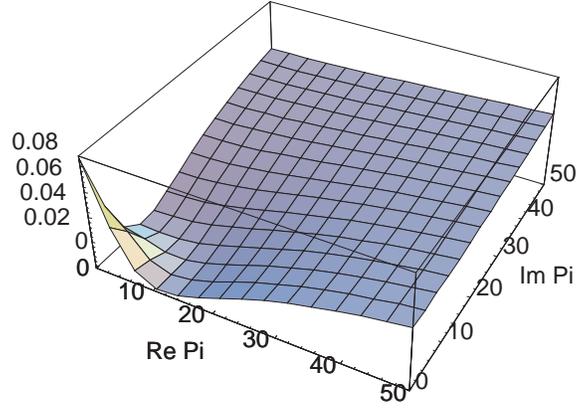

Figure 1: The connection (1.1) in the case CP(1). The value of $1/R + \Gamma^i_{kl}$ is shown. Radial directions are related with the Goldstone modes of deformations of the ellipsoid of polarizations. Angular rotations represent the Higgs modes of reorientations of this ellipsoid in every point of a geodesic. All geodesics are subjected to a "rigid rotation" under the action of the isotropy group of the "vacuum vector".

In the consistent relativistic approach one can define the density of the canonical momentum $\tilde{\mathcal{P}}(\Psi, \tau)$

$$\tilde{\mathcal{P}}(\Psi, \tau) = \frac{\delta \mathcal{L}'}{\delta(\frac{\partial \Psi}{\partial \tau})} \tag{5.3}$$

and the Hamiltonian density

$$\tilde{\mathcal{K}} = \tilde{\mathcal{P}} \frac{\partial \Psi}{\partial \tau} - \mathcal{L}' \tag{5.4}$$

in the spirit of Horwitz-Piron [12] where $\tau$ is now the parameter of the geodesic variation in CP(N).

In order to define the *dynamical shift* of the our field configuration we should introduce besides *kinematic coordinades* $x^\mu$ in the reference Minkowski spacetime *dynamical coordinates*

$$X^\mu + \delta X^\mu = < x^\mu + \delta x^\mu > = \int |\Psi + \Delta \Psi|^2 x^\mu dy. \tag{5.5}$$

Then *dynamical shift* is as followes

$$\delta X^\mu = < \delta x^\mu > = \int (|\Psi + \Delta \Psi|^2 - |\Psi|^2) x^\mu dy. \tag{5.6}$$

Then one can introduce the *Noether's current* $N^\mu$ which obeys the equation

$$\delta\mathcal{L} = \mathcal{L}' - \mathcal{L} = \frac{\delta N^\mu}{\delta X^\mu} \tag{5.7}$$

and nonperturbative version of the Ward-Takahashi identities which closely connected with dynamical symmetries and their reconstruction.

# 6 The Quantum Origin of Inertia

The problems of the physical invariance have often been connected with kinamatical changing of frame or the second copy of the identical physical system which moves relative the first one. But problem of inertia demands *dynamical* changing of a character of motion of the searching system.

It is well known that the existence and stability of extended macroscopic objects are supported by Goldstone's modes [13]. **I think we can connect the inertia property rather with the quantum conditions of excitations and reconstruction of these modes than with the Mach principle**. In some sense one can say that the source of inertia in fact lies "outside" the spatial boundary of body–just in Hilbert state space. Namely, deformation of the self-consistent boundary leads to the generation of the collective mode as a dynamical quantum reaction on external force. **It means that acceleration is only external "exibit" of internal quantum reaction [1, 2, 3]. That is one may treat the inertia as an "elasticity" of the self-consistent boundary at the quantum level**.

Here we try to connect the inertia property of an "elementary particle" and the deformation of the internal quantum collective modes which prevent this system from flying apart under some kinds of repulsive forces.

Dirac applied a surface tension force of some non-Maxwellian type as a stabilizer in the extensible model of the electron [14]. K.R.W. Jones pursuing to reach in his beautiful works [15, 16] the objective interpretation of the wave function, have used the Newtonian gravitational self-energy in order to find bound and stable solitary wave solution of the nonlinear *gravitational Schrödinger equation*. That is this nonlinearity should play the role of the physical mechanism of suppressing of the dispersion which leads to spread out wave packets. I think that for the relativistic realization of these ideas, a geometrical approach is unavoidable and that geometrical spirit of Einstein program should be conserved. **Instead of nonrelativistic scheme of Jones I propose to use the superrelativity principle which returne us to the geometrical origin (in projective Hilbert space) of the unified interactions in accordance with Einstein's program. Then the curvature $R^s_{kj^*i}$ of the CP(N) is the quantum source of the inertia of the nonlocal elementary particles**.